\documentstyle[preprint,tighten,aps,psfig]{revtex}

\begin{document}

\preprint{TTP97-10, hep-ph/9704355}
\draft

\date{April 1997}
\title{Neutron Electric Dipole Moment In The Standard Model:\\
       Valence Quark Contributions} 
\author{Andrzej Czarnecki\thanks{Address after September 1997: 
Physics Department,
Brookhaven National Laboratory,  Upton, New York 11973.} 
and Bernd Krause}
\address{Institut f\"ur Theoretische Teilchenphysik, 
Universit\"at Karlsruhe,\\
D-76128 Karlsruhe, Germany}
\maketitle

\def\be{\begin{eqnarray}}
\def\ee{\end{eqnarray}}
\def\ecm{{\mbox{ $e$ cm}}}

\begin{abstract}
We present a complete three-loop calculation of the electric dipole
moment of the $u$ and $d$ quarks in the Standard Model.  For the $d$
quark, more relevant for the experimentally important neutron electric
dipole moment, we find cancellations which lead to an order of
magnitude suppression compared with previous estimates.
\end{abstract}

\vspace*{3mm}

The electric dipole moment (EDM) of the neutron, $d_n$, has been
subject of experimental searches for almost a half century
\cite{Ramsey:1982}.  Discovery of a non-zero $d_n$ would
be direct 
evidence of violation of both time reversal symmetry and parity. 
If the symmetry under simultaneous charge conjugation, space
inversion, and time reversal (CPT) is exact, T violation is equivalent
to CP violation (see also a discussion in \cite{Kayser:1996}).  Most
theories put forward to explain CP violation observed in neutral kaon
decays, or to address the question of matter-antimatter asymmetry in
the universe, predict the existence of $d_n$ at some level.
The current experimental upper bound
\cite{RPP,Altarev:1992,Smith:1990} 
\be
d_n<1.1 \times 10^{-25} \ecm
\ee
stringently constrains physics both within the standard model
(possible non-perturbative effects, such as the so-called 
$\theta$-term \cite{Peccei}) and  beyond it, such as 
supersymmetry and supergravity \cite{Barbieri:1996},
fourth generation of fermions \cite{Hamzaoui:1995},
CP violation in the Higgs sector \cite{Kh:1996}, etc.~(for
detailed reviews see \cite{Barr:1993,BarrMarciano}).  
It also severely constrains the $W$ boson EDM \cite{MQ}.

The next generation of experiments is going to search for $d_n$ with
an accuracy improved by several orders of magnitude.  For example, a
new technique based on storing polarized ultra cold neutrons and a
polarized gas of $^3$He in superfluid $^4$He is expected to permit a
measurement of $d_n$ with the sensitivity of 
\be
\sigma(d_n)=4\times 10^{-29} \ecm
\ee
in a one year run \cite{Golub:1994}. 

In the electroweak sector of the standard model a complex phase of the
Cabibbo-Kobayashi-Maskawa (CKM) matrix violates CP and,  via loop
effects,  induces an EDM for all non-selfconjugate particles with spin. 
In the case of the neutron, an EDM can be generated by electroweak
interactions in various ways. 
Penguin diagrams with a $W$ exchange between
constituent quarks  offer one possibility.  Such effects were estimated to 
give contributions to
$d_n$ which could perhaps be as large as 
$10^{-30} \ecm$ \cite{Gavela:1982}. 
This estimate is controversial and involves a large uncertainty; see
e.g.~a discussion in \cite{Bigi:1991a}.  A more conservative
estimate of the long distance standard model effects is $2\times
10^{-32} \ecm$ \cite{KhZ}.
 
\begin{figure}
\hspace*{2mm}
\begin{minipage}{25.cm}
\[
\mbox{
\hspace*{-67mm}
\begin{tabular}{cc}
\psfig{figure=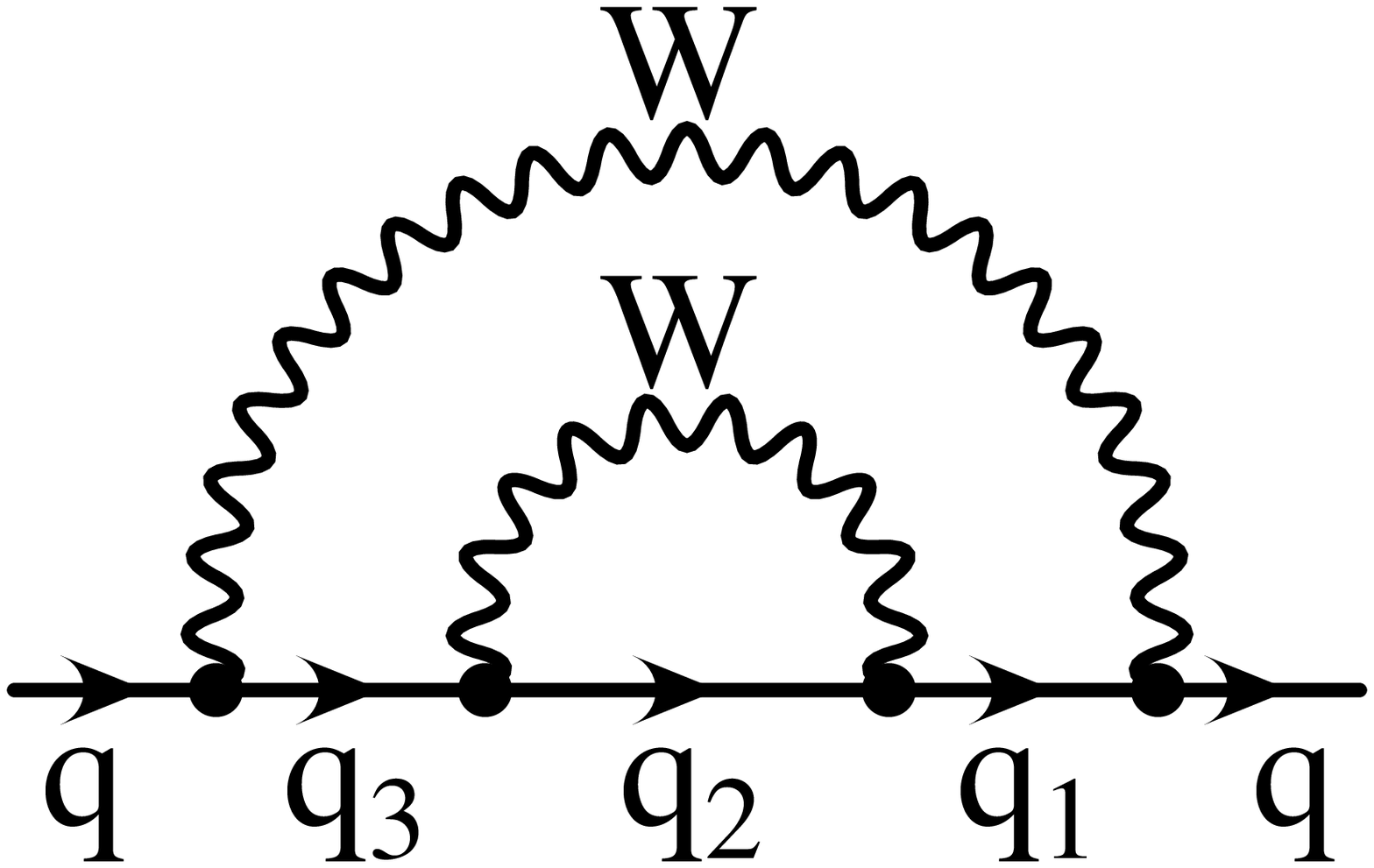,width=35mm,bbllx=210pt,%
bblly=600pt,bburx=630pt,bbury=550pt} 
&\hspace*{.6cm}
\psfig{figure=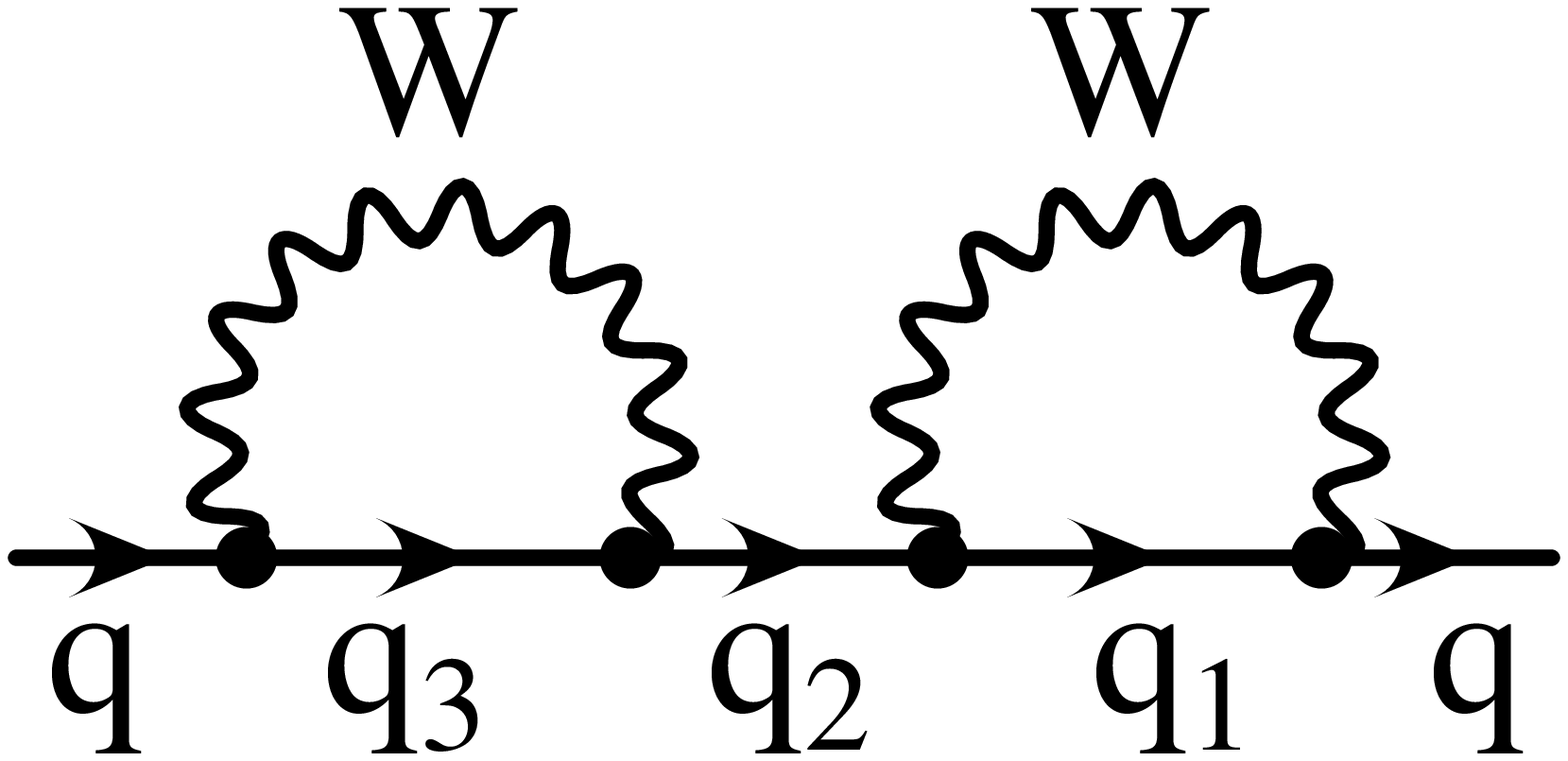,width=35mm,bbllx=210pt,%
bblly=600pt,bburx=630pt,bbury=550pt}
\\[22mm]
\rule{-18mm}{0mm} (a) &\hspace*{-.4cm} \rule{-6mm}{0mm}(b)
\end{tabular}}
\]
\end{minipage}
\caption{Flavor structure of diagrams contributing to quark EDM.
  Gluon lines are not indicated.  Here and in the following figures it
  is understood that an external electric field can interact with any
  charged particle in the diagram.}
\label{fig:flavor}
\end{figure}

Another
possibility, and the main focus of our work, is $d_n$
induced by an EDM of the up and down quarks ($d_{u,d}$) 
inside the neutron.  Using SU(6)
wave functions one finds
\be
d_n = {4\over 3}d_d -{1\over 3} d_u.
\ee
Quark EDM cannot be generated in the standard model
at the one loop level because the relevant amplitudes do
not change the quark flavor and each CKM matrix element is accompanied
by its complex conjugate; no T-violating 
complex phase can arise.  At the two loop
level individual diagrams have complex phases (see
Fig.~\ref{fig:flavor})  
and contribute to the EDM
\cite{Maiani:1976,Ellis:1976,Lee:1977}.  However, the sum over all
quark flavors in the intermediate states leads to the vanishing of the
EDM
at two loops \cite{Shabalin:1978} (a similar effect for leptons was
found in ref.~\cite{Donoghue:1978}).  This interesting and seemingly
accidental cancellation was also analyzed in 
\cite{Khriplovich:1986R,Czar:1996dip}.  

The fact that the quark EDM appears only at the three loop level in
the standard model greatly complicates theoretical estimates.  The
largest effect is due to the exchange of two $W$ bosons and one gluon;
this was first discussed by Shabalin in \cite{Shabalin:1980} in a
model with quarks carrying integral charges 0 and 1.  In that model
the charged quark EDM was found to be suppressed by at least 6 inverse
powers of the $W$ mass, if one assumes quarks to be light.  The case
of quarks with physical, fractional values of electric charges was
examined in \cite{Khriplovich:1986R}.  It was found
that the strong cancellation found earlier was an artifact of the
assumed pattern of charges and quark EDM was found at the level of
(quark masses)$^4/M_W^4$ with further enhancements by logs of
quark mass ratios; in 
fact, in this order the EDM of the quark is proportional to the
electric charge of its isospin partner --- so that the EDM of a charge
1 quark in Shabalin's model is suppressed.

The analysis of ref.~\cite{Khriplovich:1986R}, the
most complete study to date, was carried out in the framework of the
leading logarithmic approximation within the
effective Fermi theory under an assumption that all quarks are lighter
than the $W$ boson.  Since it turns out that the top quark outweighs
the $W$ boson, this approach is no longer valid.  In fact, the large
mass of the top quark significantly enhances some electroweak
processes (e.g.~the decays $Z\to b\bar b$ and $b\to s\gamma$, as well
as $B^0-\bar B^0$ mixing).  In view of the anticipated improvement of
the experimental accuracy a detailed evaluation of the standard model
contributions to the quark EDM is clearly warranted.  Here, we report the
complete results of such an analysis.

The quark matrix element of the electromagnetic current,
$J_\mu^{\rm em}$, can be written as (with $q=p'-p$)
\begin{eqnarray}
\lefteqn{\!\!\!\!\!\!\!
\langle p'| J_\mu^{\rm em} | p \rangle = \bar u(p')
\Gamma_\mu u(p)} \nonumber \\
\Gamma_\mu &=& 
 F_1(q^2)\gamma_\mu 
+ i F_2(q^2) \sigma_{\mu\nu} q^\nu
- F_3(q^2)\gamma_5 \sigma_{\mu\nu} q^\nu
\nonumber\\
&&+F_A(q^2) \left( \gamma_\mu q^2 - 2m_q q_\mu\right) \gamma_5
\label{eq:emcurrent}
\end{eqnarray}
The form factors at $q^2=0$ give the electric charge,
anomalous magnetic moment, electric dipole moment, and anapole moment 
in units of
$e=|e|$ (e.g.~for a down quark $F_1(0)=-1/3$), so that $d_q = eF_3(0)$.

The three-loop calculation of the light quark EDM is facilitated by the
clear hierarchy of masses $m_{u,d}\ll m_s \ll m_c \ll m_b \ll M_W \ll
m_t$.  On the other hand, as will be seen shortly, the relevant 
mass ratios are
not large enough for their logarithms to be the dominant effect. 
  Therefore, in addition to the logarithmic terms we also need
the non-logarithmic constants. 
A systematic expansion of Feynman diagrams in terms of mass ratios and
their logs is possible in the framework of asymptotic expansions
(for a recent review see \cite{Smi94}).  The problem is slightly
complicated by the many mass scales present in each diagram.  The
details of our procedures will be described elsewhere.  Here we only
remark that the calculation technically resembles the two-loop
electroweak corrections to the muon anomalous magnetic moment
\cite{CKM95}, apart from the presence of the third loop.

\begin{figure}
\hspace*{2mm}
\begin{minipage}{25.cm}
\[
\mbox{
\hspace*{-67mm}
\begin{tabular}{cc}
\psfig{figure=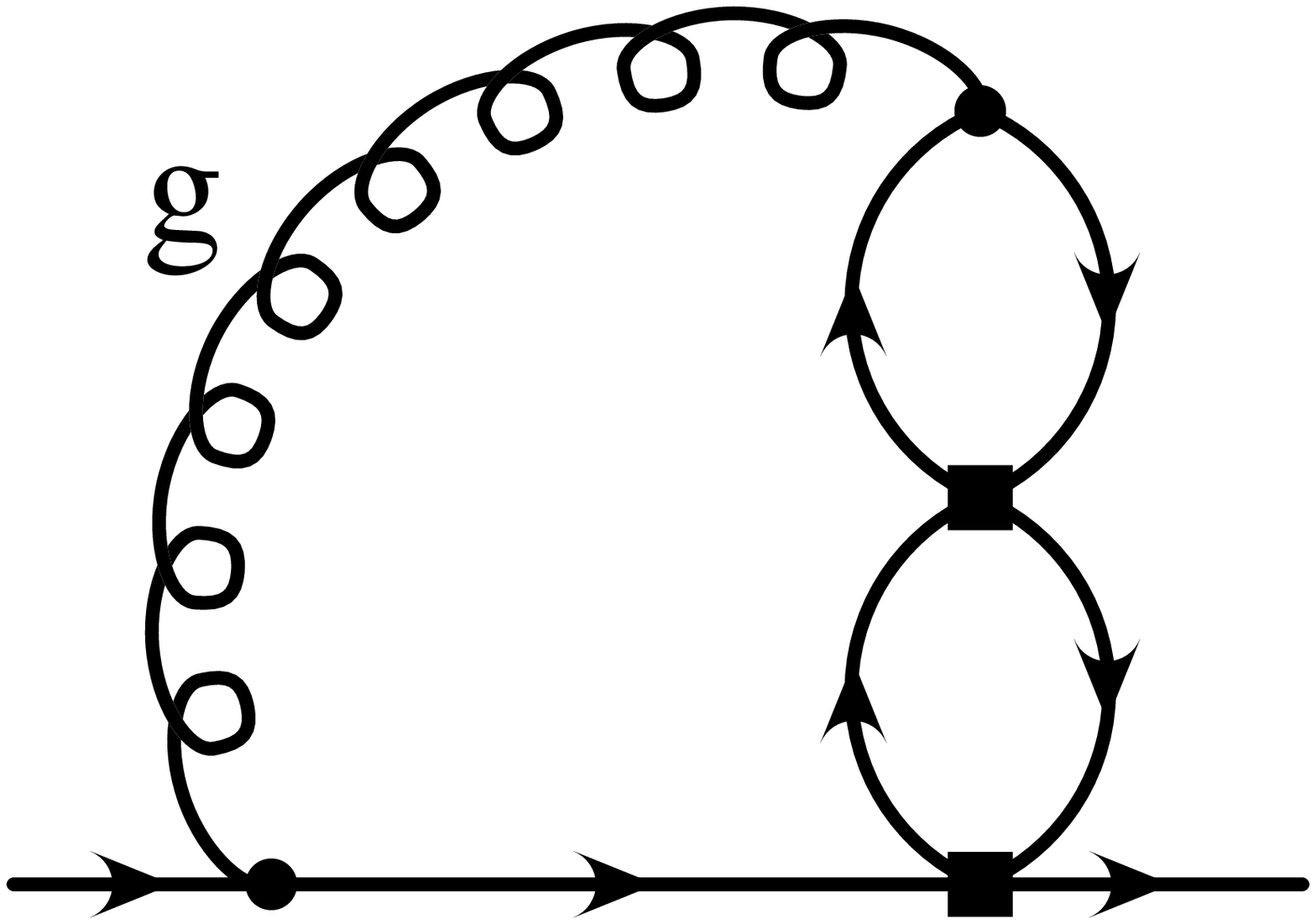,width=35mm,bbllx=210pt,%
bblly=660pt,bburx=630pt,bbury=350pt} 
&\hspace*{.6cm}
\psfig{figure=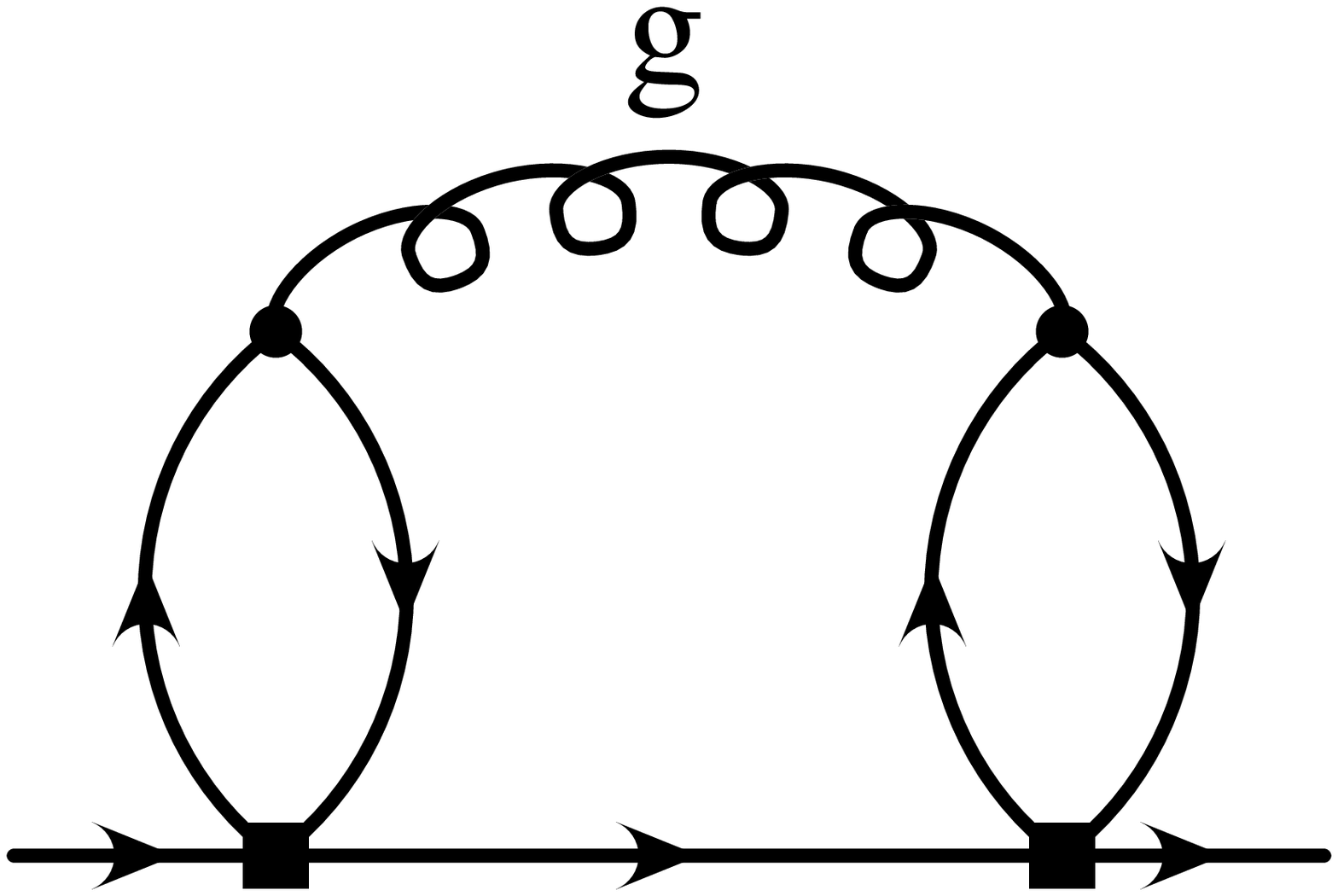,width=35mm,bbllx=210pt,%
bblly=660pt,bburx=630pt,bbury=350pt}
\\[26mm]
\rule{-18mm}{0mm} (a) &\hspace*{-.4cm} \rule{-6mm}{0mm}(b)
\end{tabular}}
\]
\end{minipage}
\caption{Effective Fermi theory diagrams in a model with all quarks
  lighter than the $W$ boson.}
\label{fig:Fermi}
\end{figure}

The electric dipole moment in the standard model has an important
feature which simplifies its calculation.  Namely, if the source of CP
violation is in the complex phase of the CKM matrix, the EDM vanishes if
any two up-type or down-type quarks have equal masses.  This
circumstance has been taken advantage of in the Fermi theory
calculation in ref.~\cite{Khriplovich:1986R}. Under
the assumption that all quarks are lighter than the $W$ boson it was
possible to replace the $W$ propagators by $1/M_W^2$; effects of large
virtual momenta which could ``feel'' the exact structure of these
propagators are, to first approximation, independent of quark masses,
and therefore their sum gives no contribution to the EDM.  The resulting
effective theory diagrams are shown in Fig.~\ref{fig:Fermi}.

\begin{figure}
\vspace*{-5mm}
\hspace*{2mm}
\begin{minipage}{25.cm}
\[
\mbox{
\hspace*{-67mm}
\begin{tabular}{c}
\psfig{figure=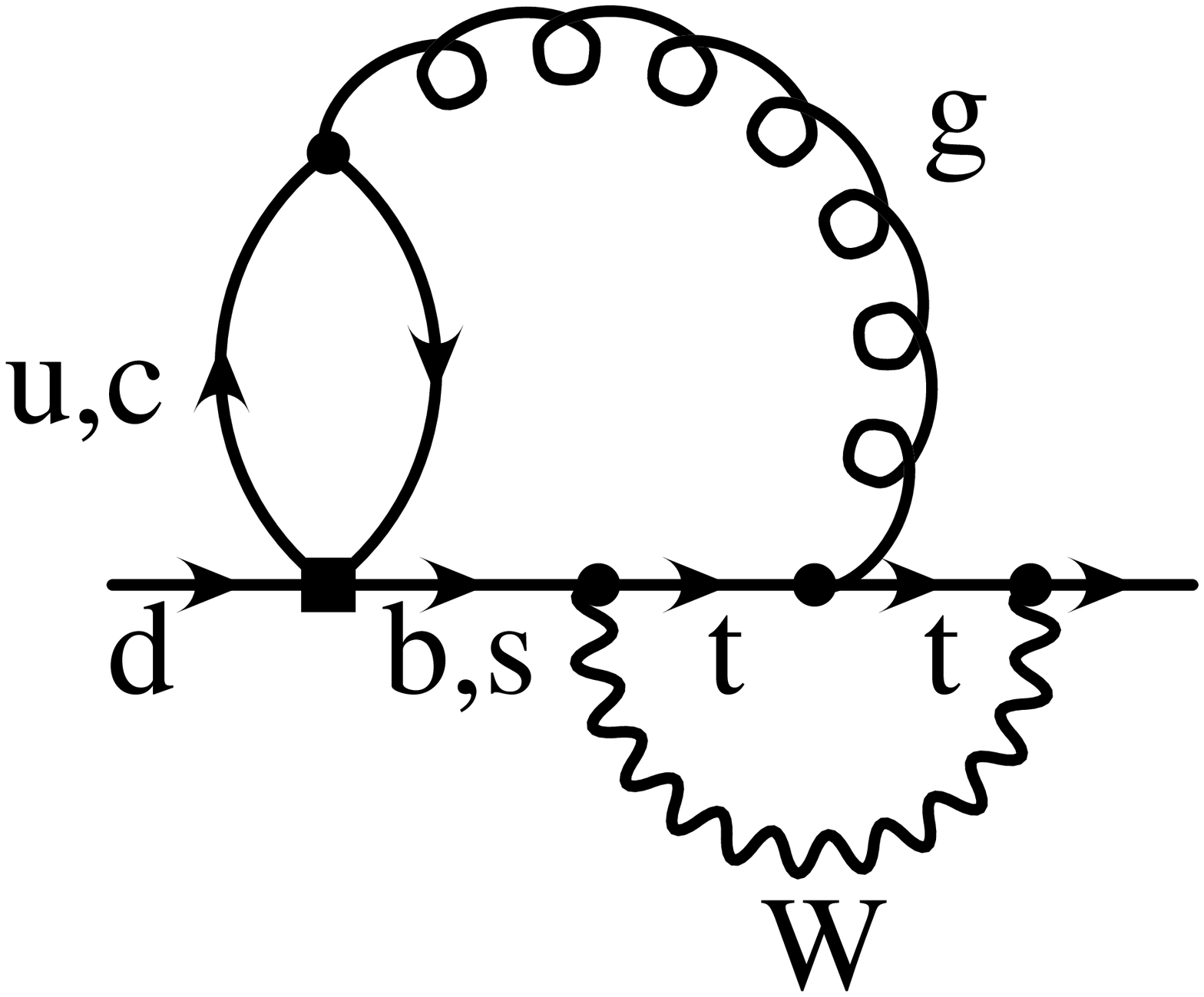,width=32mm,bbllx=210pt,%
bblly=660pt,bburx=630pt,bbury=350pt} 
\end{tabular}}
\]
\vspace*{23mm}
\[
\mbox{
\hspace*{-70mm}
\begin{tabular}{cc}
\psfig{figure=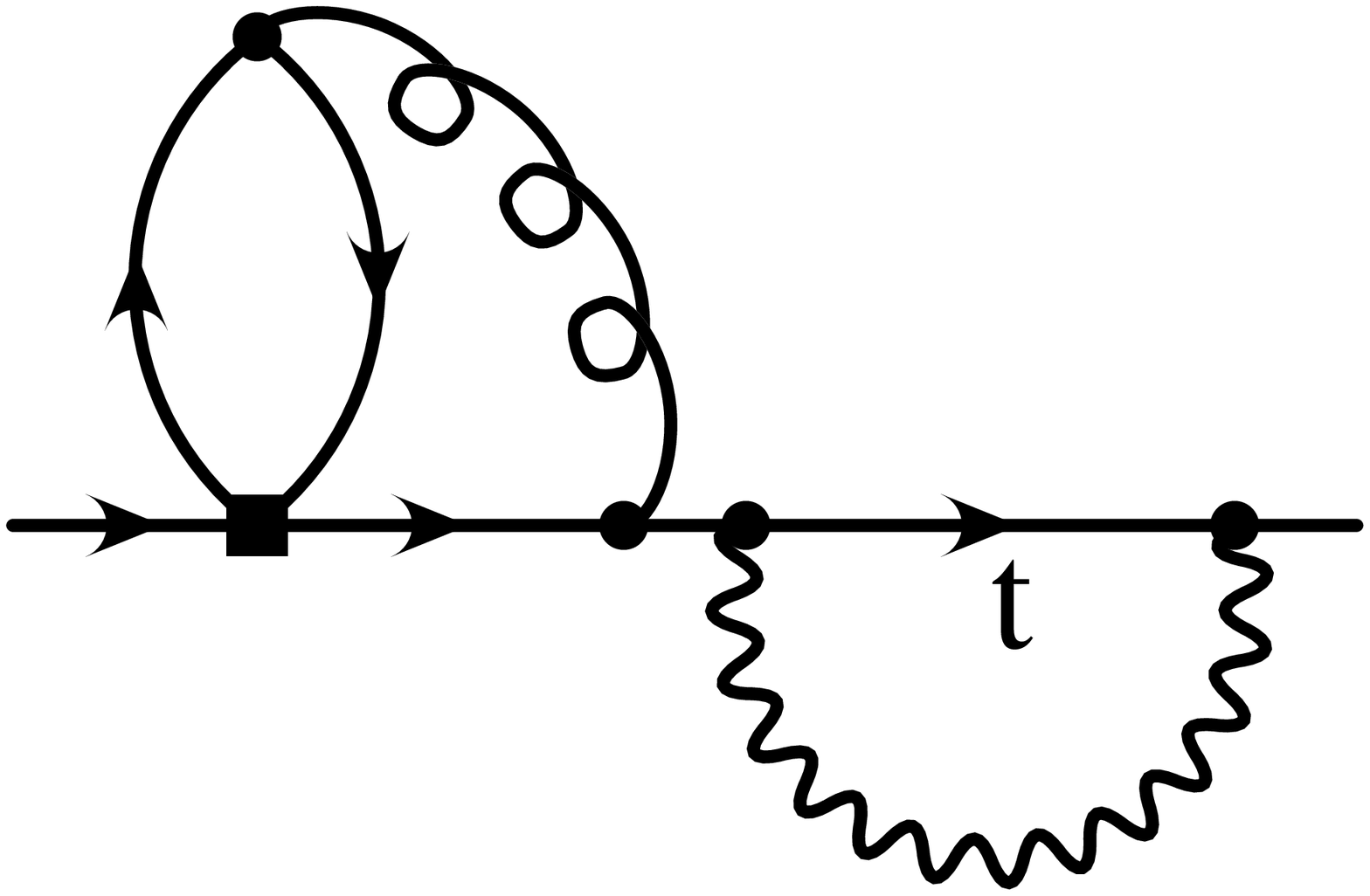,width=32mm,bbllx=210pt,%
bblly=660pt,bburx=630pt,bbury=350pt} 
&\hspace*{8mm}
\psfig{figure=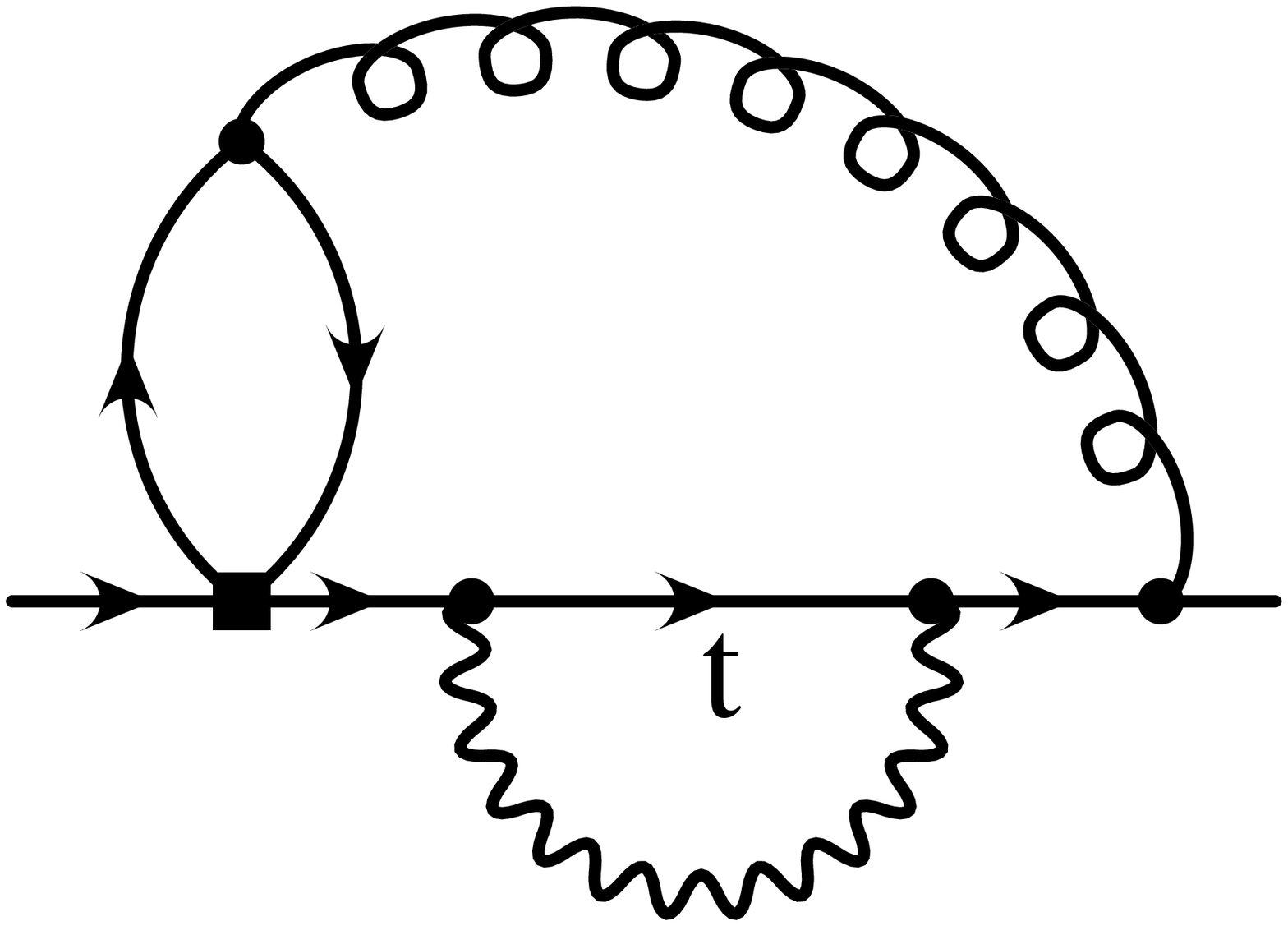,width=32mm,bbllx=210pt,%
bblly=660pt,bburx=630pt,bbury=350pt}
\\[29mm]
\end{tabular}}
\]
\end{minipage}
\caption{Heavy top quark contributions to $d_d$.}
\label{fig:dd}
\end{figure}

\begin{figure}
\hspace*{2mm}
\begin{minipage}{25.cm}
\[
\mbox{
\hspace*{-67mm}
\begin{tabular}{c}
\psfig{figure=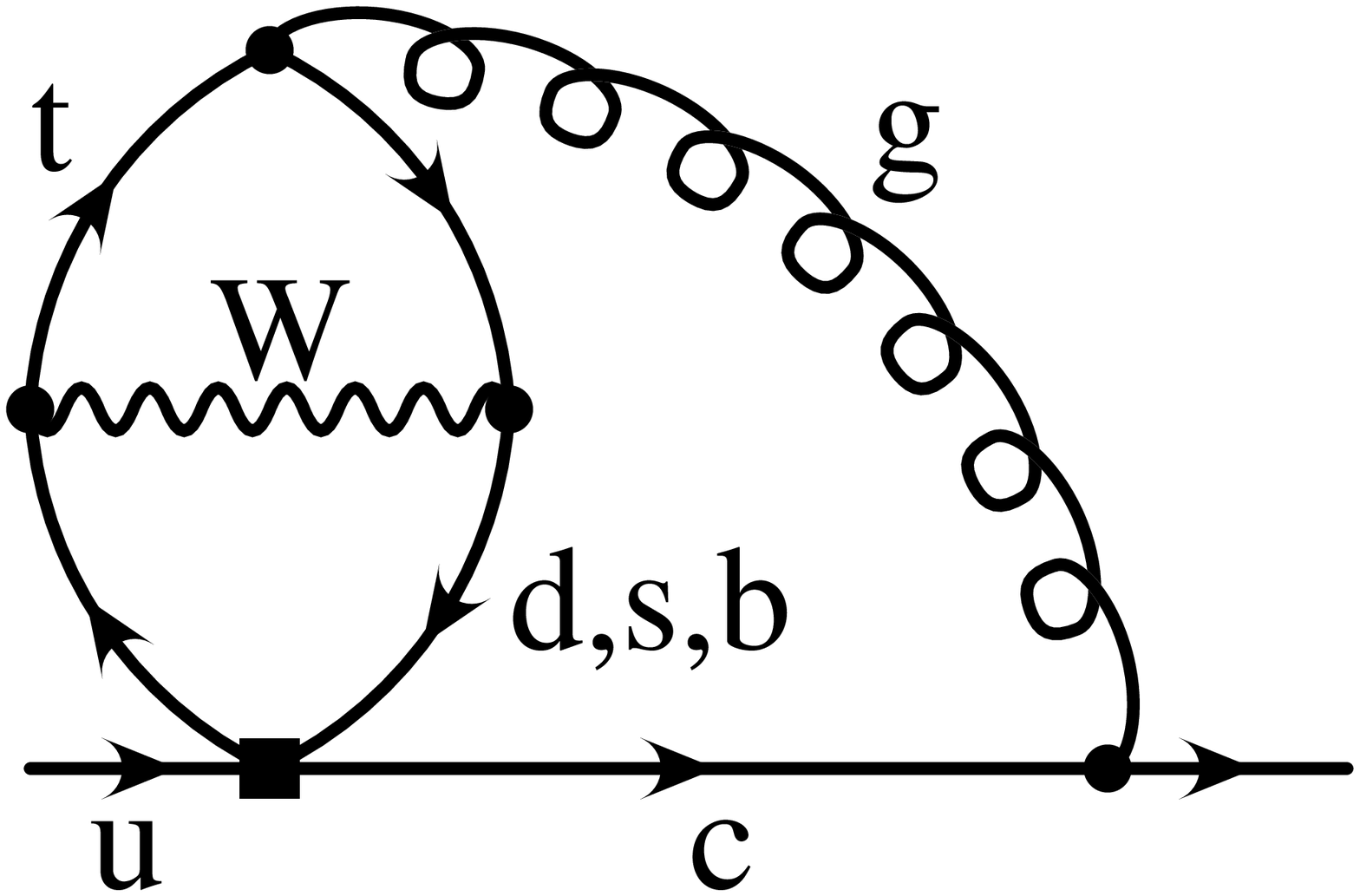,width=32mm,bbllx=210pt,%
bblly=660pt,bburx=630pt,bbury=350pt} 
\end{tabular}}
\]
\vspace*{13mm}
\[
\mbox{
\hspace*{-70mm}
\begin{tabular}{cc}
\psfig{figure=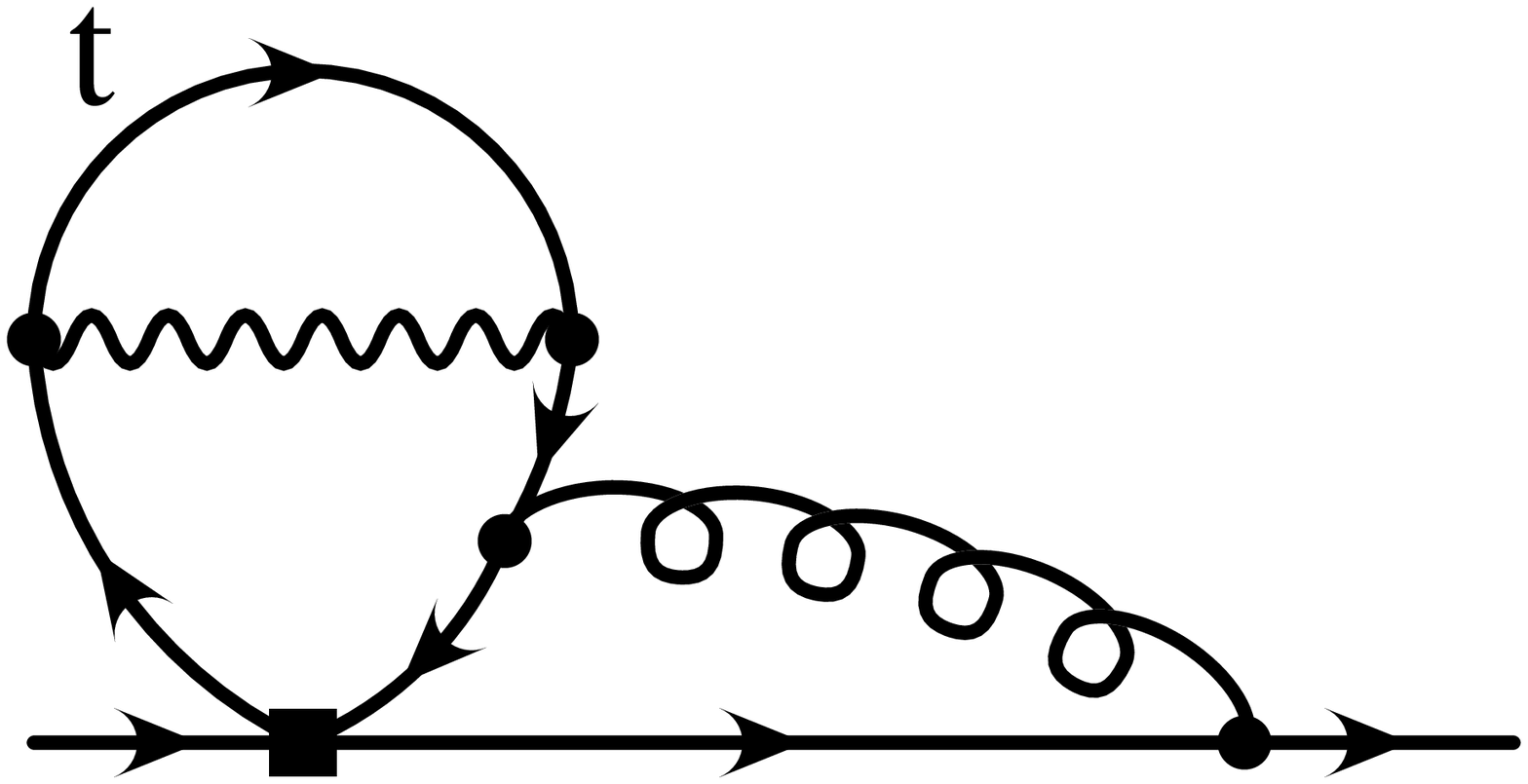,width=32mm,bbllx=210pt,%
bblly=660pt,bburx=630pt,bbury=350pt} 
&\hspace*{8mm}
\psfig{figure=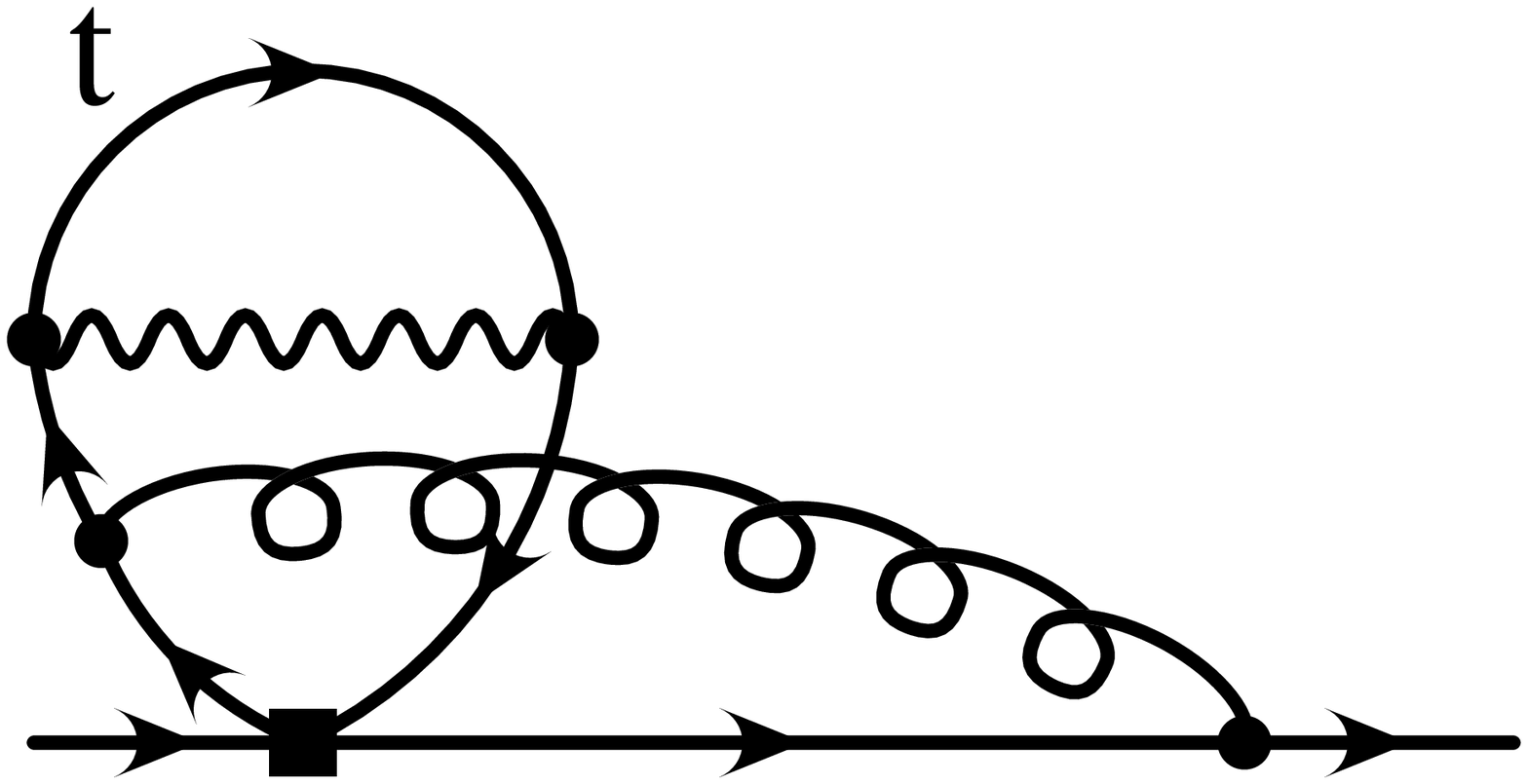,width=32mm,bbllx=210pt,%
bblly=660pt,bburx=630pt,bbury=350pt}
\\[23mm]
\end{tabular}}
\]
\end{minipage}

\caption{Heavy top quark contributions to $d_u$.}
\label{fig:du}
\end{figure}

The situation is actually more complicated since the top quark is
heavier than the $W$ boson.  It turns out that there are 6 topologies
of diagrams with the top quark which contribute to the EDM: 3 for the down
quark (Fig.~\ref{fig:dd}) and 3 different ones for the up quark
(Fig.~\ref{fig:du}).  The diagrams in Figs.~\ref{fig:dd} and
\ref{fig:du} have to be understood in the following way: the $W$
propagator connected to light quark lines only can be contracted
to a point and replaced by $1/M_W^2$; the other $W$ propagator,
connected to the top quark and indicated explicitly in the figures, is
treated differently.  Namely, the entire top-$W$ loop is replaced by
an effective operator insertion.  Technically this amounts to
performing the integral over its obvious internal momentum, after
having expanded top and $W$ propagators in remaining momenta and
$M_W$.  We note that contributions of internal momentum in the $t-W$
loop of the order much less than $m_t$ are suppressed; this is obvious
in the unitary gauge which we use in this calculation. 

Contributions of diagrams containing the top quark are finite and
independent of $m_t$ for both up and down type quarks (throughout
this paper we adopt the approximation in which we neglect terms
suppressed by inverse powers of quark masses).  In particular, we have
found no terms enhanced by $m_t^2$, like $m_t^2/M_W^6$. 

The question which now arises is whether the effective Fermi
theory is sufficient for the 
calculation of  the
remaining diagrams, without the top quark.  
This turns out not to be the case.  In diagrams with an external down
quark we have to include the contributions depicted in
Fig.~\ref{fig:dd} with the top quark replaced by up and charm quarks.
Similarly, for an external up quark, diagrams like in
Fig.~\ref{fig:du} contribute, with $t\to c$.  In these diagrams the internal
momenta of the order of $M_W^2$ give contributions to the EDM which are no
longer cancelled by the corresponding top diagrams.   
They have to be added to the (divergent)
effective Fermi theory results.  In the sum divergences cancel and
logarithms of $M_W$ are combined with those of the mass of the second
heaviest quark, $m_b$. 
In this way we arrive at our final
formulas for the down quark
\be
\lefteqn{   {d_d\over e} =
 {m_d m_c^2 \alpha_s G_F^2   \tilde\delta\over 108\pi^5} 
\left[
 \left(       L ^2 _{bc }
       -2 L_ {bc }
       +  {\pi^2\over 3}\right)  L_ {Wb } \right.}
 \nonumber \\ && \left.
       +  {5 \over 8}  L ^2 _{bc } 
       - \left(   {335 \over 36} + {2 \over 3} \pi^2 \right) L_ {bc } 
       - {1231 \over 108}+ {7 \over 8} \pi^2 + 8 \zeta_3
\right] +{\cal O}(m^2/M^2),
\ee
and for the up quark:
\be
{d_u\over e} &=&
 {m_u m_s^2 \alpha_s G_F^2   \tilde\delta\over 216\pi^5} 
\left[
 \left( - L ^2 _{bs } +2 L_ {bs } + 2 - {2\pi^2\over 3} \right)L_ {Wb} 
\right.
\nonumber \\ &&
       -L_ {bc } L ^2 _{cs } 
      + 2L_ {bc } L_ {cs } 
       -{5 \over 8} L ^2 _{bs } 
       - \left(   {259 \over 36} + {\pi^2 \over 3}  \right) L_ {bs } 
 \nonumber \\ && \left.
       + \left( {140 \over 9} + \pi^2 \right) L_ {cs } 
       - {121 \over 108} + {41 \over 36} \pi^2 - 4 \zeta_3
\right]+{\cal O}(m^2/M^2),
\ee
where ${\cal O}(m^2/M^2)$ denotes terms with an additional suppression
       by quark or $W$ boson masses;
$L_{ab}\equiv \ln(m_a^2/m_b^2)$, $\zeta_3$ is the Riemann zeta
function ($\zeta_3=1.202\ldots$), and $\tilde\delta$ denotes the CP
violating invariant
\be
\tilde\delta = s_1^2s_2s_3c_1c_2c_3\delta
\ee
where for the CKM matrix we use the convention of
\cite{Khriplovich:1986R,Okun81}.  
From our expressions it is not obvious that the quark EDM vanishes
if any two up- or down-type quarks have degenerate masses.  This is
because we have assumed a hierarchy of masses, and retained only the
leading terms in the expansions in mass ratios.  Such expansion, while
delivering a simple and compact formula, obscures the symmetry
of the full result.

It is instructive to compare our findings with the results obtained in
the Fermi theory \cite{Khriplovich:1986R}.  In that
reference the terms with three powers of logarithms were calculated
and if the top quark mass is replaced by mass of the $W$ boson they
coincide with our $L_{bc}^2L_{Wb}$ (for $d_d$) and $-L_{bs}^2L_{Wb} -
L_{bc}L_{cs}^2$ (for $d_u$).  However, it turns out that the terms
with the highest power of logarithms are not the dominant components
of the complete result. In fact, the large negative coefficient of the
single logarithms $L_{bc}$ leads to more than a complete cancellation
of the triple log contribution to $d_d$.  Partial cancellation is also
found in $d_u$.  For the  numerical estimates  we use the following
values of parameters: $\tilde\delta=5\times 10^{-5}$, $\alpha_s=0.2$,
$m_s=0.2$ GeV, $m_c=1.5$ GeV, $m_b=4.5$ GeV, $M_W=80$ GeV.   We find
\be
d_d &=& -0.7\times 10^{-32} {m_d\over \mbox{GeV}} \ecm
\nonumber \\
d_u &=& -0.3\times 10^{-32} {m_u\over \mbox{GeV}} \ecm
\label{eq:num_result}
\ee
Previous estimates were based only on 
the terms with three powers of logs; from those we find
\be
d_d(\mbox{triple log approx.}) &=& +\,5\hspace{2.5mm} \times  10^{-32}
{m_d\over 
  \mbox{GeV}} \ecm 
\nonumber \\
d_u(\mbox{triple log approx.}) &=& -0.4  \times  10^{-32} {m_u\over
  \mbox{GeV}} \ecm 
\ee
We see that the ``non--leading-log'' terms
suppress the quark EDM, especially  strongly for the $d$ quark where they
even change the sign of the effect.  For the
experimentally interesting neutron EDM, the $d$ quark EDM is the more
important quantity. The suppression factor we found decreases the part
of the standard model contribution generated by the quark EDM and renders
it definitely unobservable for the experiments in the near future. 

Inserting current quark masses for $m_{u,d}$ in
eq.(\ref{eq:num_result}) we obtain the following numerical values 
\be
d_d &=& -0.7\times 10^{-34} \ecm \quad\mbox{for $m_d=10$ MeV}
\nonumber \\
d_u &=&  -0.15\times 10^{-34} \ecm \quad\mbox{for $m_u=5$ MeV}
\ee


We should stress here that we have neglected terms with more than one
power of light quark masses $m_{u,d}$.  They tend to result in an
additional  suppression, especially for $d_u$, where
$m_s^2$ should be replaced by $m_s^2-m_d^2$ in most terms. 

 Regarding
the accuracy of our result, it is clear that the issue of what light 
quark masses should be used is  most important and requires further
study.  
This is connected with the issue of the scale of the running
$\alpha_s$ which is difficult to address since our calculation is of
the leading order in QCD. For the triple-logarithmic term it is known
from \cite{Khriplovich:1986R} that the characteristic gluon virtuality
is between $c$ and $b$ quark masses (for $d_d$). 
For the non-leading terms this scale could be different.
Also, since $d_d$ results from large cancellations among terms
which are 5 to 10 times larger than the final value, a precise
prediction would require taking into account corrections of the order
$M_W^2/m_t^2\approx 20\%$ and $m_c^2/m_b^2\approx 10\%$
which can change the result by a factor of 2
or so.  It is, however, rather certain that the non--leading-log terms
we presented in this paper significantly reduce the value of $d_d$
compared with previous estimates. If the next
generation of experiments detects non-zero $d_n$ it will have to be
attributed to a different mechanism than the standard model light quark
EDM.

\subsection*{Acknowledgment}

We are very grateful to Professors I.~Khriplovich, F.~Klinkhamer,
W.~Marciano, and N.~Uraltsev for advice and helpful comments.  We
thank Professor J.~H.~K\"uhn for his interest in this work and
support.  This~research was supported by the grant BMBF 057KA92P and
by ``Gra\-duier\-ten\-kolleg Elementar\-teil\-chen\-phy\-sik'' at the
University of Karlsruhe.


\end{document}